\documentclass[letterpaper]{article}
\usepackage[preprint]{aaai2027}
\nocopyright
\newcommand{\methodname}{CLASVS}
\usepackage[hyphens]{url} 
\usepackage{graphicx} 
\usepackage{natbib} 
\usepackage{caption} 
\usepackage{booktabs}
\usepackage{amsmath}
\usepackage{amssymb}

\title{\methodname{}: Continuous-Latent Autoregression for Melody-Preserving \\ Lyric Editing in Singing Voice Synthesis}
\author{Yizhong Geng\textsuperscript{\rm 1,2}\thanks{Equal contribution.}, Tian-Hao Zhang\textsuperscript{\rm 2,3}\footnotemark[1], Chunfeng Wang\textsuperscript{\rm 2}, Wenxin Fu\textsuperscript{\rm 1}, Yingming Gao\textsuperscript{\rm 1},\\
Ruimin Wang\textsuperscript{\rm 2}, Zhou Pan\textsuperscript{\rm 2}, Kun Zhan\textsuperscript{\rm 2}\thanks{Corresponding authors.}, Liang Li\textsuperscript{\rm 3}, Ya Li\textsuperscript{\rm 1}\footnotemark[2]}
\affiliations{\textsuperscript{\rm 1}Beijing University of Posts and Telecommunications, Beijing, China\\
\textsuperscript{\rm 2}Li Auto Inc., Beijing, China\\
\textsuperscript{\rm 3}Tsinghua University, Beijing, China}

\begin{document}

\maketitle

\begin{abstract}
Reference-conditioned melody-preserving lyric editing replaces words while retaining a performance's timing, singer identity, and naturalness. Continuous-latent autoregression avoids finite codebooks and offers stepwise generation with learned stopping. Editing creates a conflict absent from ordinary reconstruction: training pairs reference cues with original lyrics, whereas inference asks revised lyrics to override source-lyric-correlated cues; one source-following patch can propagate through AR history. We introduce \methodname{}. Its State--Control--Transition (SCT) routing keeps target-lyric and reference-melody controls persistent, returns semantic feedback on phonetic progress to the causal planner, and confines the previous latent patch to the local Transition. Progressive State--Control Grounding (PSCG) learns this contract through paired-edit-free, content-consistent Mandarin reconstruction. On two Mandarin benchmarks, \methodname{} improves all four operations over discrete-AR Vevo2 and reduces macro-PER by 46.2\%, while maintaining melody, singer similarity, and perceptual quality. Together, these results establish a strong continuous-AR operating point for score-annotation-free lyric edits and a basis for broader stepwise control. Audio demonstrations are available on our project page: \url{https://piedpiperg.github.io/clasvs-demo/}.
\end{abstract}


\section{Introduction}

Melody-preserving lyric editing changes the words of a singing performance while retaining its melody, timing, singer identity, and naturalness. Score-conditioned singing voice synthesis provides precise control through aligned phonemes, notes, and durations, but such annotations are costly for arbitrary recordings \cite{DiffSinger,VISinger2,XiaoiceSing2,PeriodSinger,PromptSinger,TCSinger,OpenCpop,M4Singer,GTSinger,bai2026hq}. Reference-conditioned lyric editing instead accepts a singing recording and revised lyrics directly \cite{EditSinger,Vevo2,YingMusicSinger,MeloDISinger}. Removing score annotations, however, creates a selective-preservation problem: the model must retain the performance characteristics carried by the reference while rejecting its source lyrics.

Two architectural axes shape selective preservation. Discrete codecs compact acoustic streams for autoregressive (AR) modeling \cite{SoundStream,EnCodec,DAC,TokSing,Vevo2}, but finite codebooks may lose subtle variation, whereas continuous or tokenizer free latent models reduce dependence on fixed codebooks and offer strong fidelity or scalable generation \cite{Voicebox,NaturalSpeech2,MatchaTTS,MELLE,ContinuousAudioLM}. Generation order adds a complementary tradeoff: continuous nonautoregressive (NAR) systems such as YingMusic-Singer-Plus (YingMusic+) enable parallel generation, whereas AR provides stepwise state and learned stopping. Existing continuous AR models such as DiTAR and SemaVoice mainly target content consistent synthesis \cite{DiTAR,SemaVoice}, leaving unresolved the supervision mismatch in reference-conditioned lyric editing under reconstruction only supervision.

This mismatch is especially severe in singing. Recent continuous or tokenizer free AR speech systems learn content consistent synthesis \cite{dotstts,VibeVoice,VoxCPM}, while unified speech editors rely on explicit instruction based editing supervision \cite{MingUniAudio}; neither setting directly provides singing counterfactuals, because resinging different lyrics also changes timing, phrasing, and acoustics. We therefore train on same-recording tuples whose reference, transcript, and acoustic target come from one performance, but this scalable reconstruction never exposes a conflict between reference and target lyrics or an edit template. At inference, revised lyrics create this unseen conflict, and one patch that follows the source lyrics can bias later AR history, requiring generalization without paired edits rather than memorized edit transformations.

\methodname{} addresses this challenge with a State--Control--Transition (SCT) editing contract. Target lyrics and reference melody remain persistent \emph{Control}; a frozen causal semantic encoder returns feedback on realized phonetic progress as recurrent \emph{State} \cite{ContentVec,WavLM,SVPT}; and the previous AudioVAE latent patch enters only the local \emph{Transition}. This separates global lyric planning from local acoustic continuity. Latent patches never re-enter the causal planner, preserving continuity without biasing lyric planning. Progressive State--Control Grounding (PSCG) grounds this routing through paired-edit-free, content-consistent reconstruction. State Grounding pretrains then freezes the semantic encoder, Control Grounding perturbs melody cues and the previous latent patch, and Task Grounding transfers Mandarin coverage to curated singing. It needs no paired counterfactual edit recordings.

We evaluate two Mandarin benchmarks under a common automatic-input protocol using reference audio and target lyrics. Vevo2 is the primary comparator because it shares \methodname{}'s sequential interface but predicts discrete acoustic tokens rather than continuous AudioVAE latent patches \cite{Vevo2}. \methodname{} improves all four operations over Vevo2, reducing macro-PER from .0699 to .0376 (46.2\%); the profile transfers to the public benchmark. YingMusic+ is the complementary continuous-NAR comparator \cite{YingMusicSinger}: it remains stronger on substitution, global duration, and steady-tail speed, while Vevo2 has the lowest cold-start RTF; \methodname{} leads deletion, insertion, macro-PER, voiced-onset MAE, naturalness, and lyric intelligibility. The two comparators expose complementary system trade-offs. Matched ablations provide within-stack SCT/PSCG evidence. Our contributions are:
\begin{itemize}
    \item \textbf{SCT Editing Contract.} We formulate editing as a conflict among persistent target-lyric and reference-melody controls, phonetic progress, and the previous latent patch, and route them separately.
    \item \textbf{Paired-Edit-Free PSCG.} We learn this SCT routing through three-stage, content-consistent Mandarin reconstruction, without paired counterfactual edit recordings.
    \item \textbf{Continuous-AR Editing Results.} Under the common automatic-input protocol, \methodname{} reduces macro-PER by 46.2\% versus discrete-AR Vevo2 across two Mandarin benchmarks, improving all four operations while preserving melody, similarity, and perceptual quality.
\end{itemize}

\begin{figure*}[t]
    \centering
    \includegraphics[width=\textwidth]{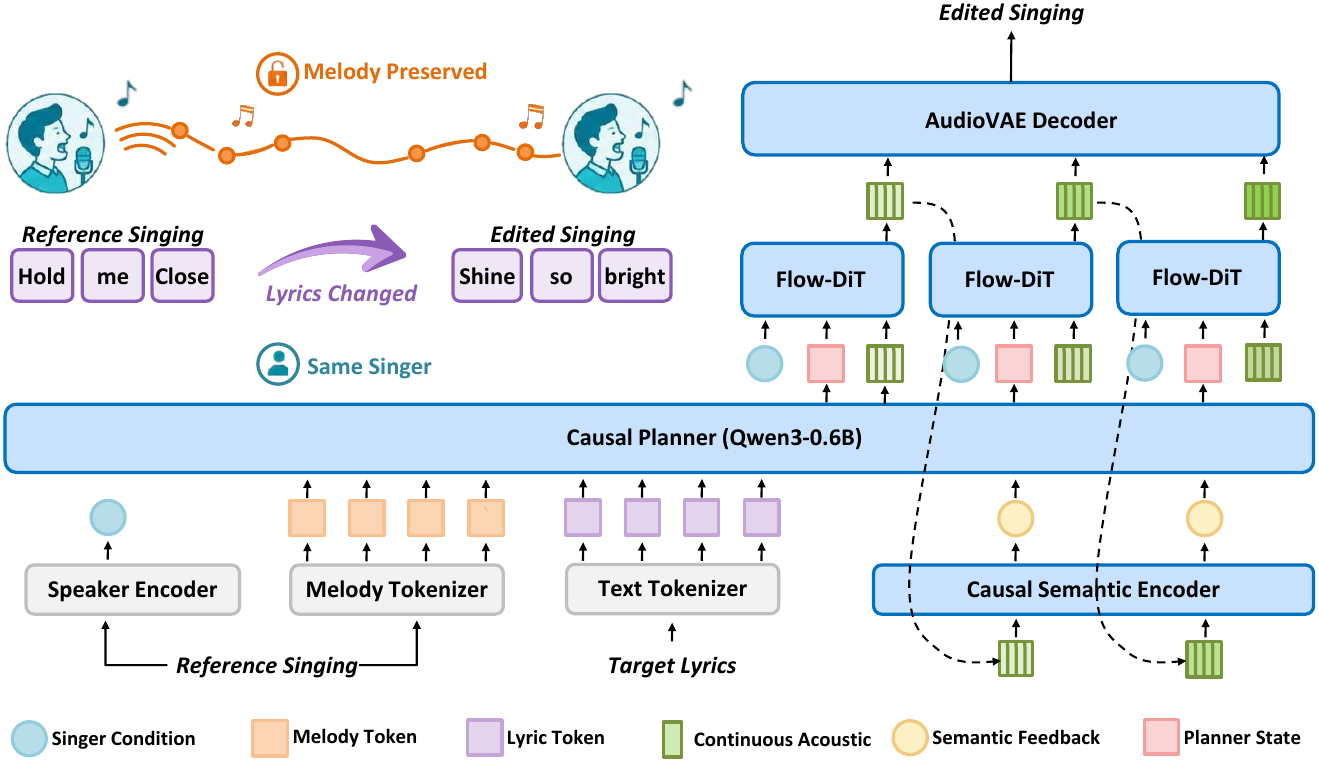}
    \caption{\methodname{} generates 64-D continuous AudioVAE latent patches every 100~ms. Target lyrics and reference-melody tokens persist in the causal planner's cache. Each patch updates semantic feedback and becomes the previous latent patch for the next step; the concatenated sequence is decoded once. Repeated Flow-DiT blocks share parameters.}
    \label{fig:sct_overview}
\end{figure*}

\section{Method}

\subsection{Inputs, Time Bases, and Duration}

Let $\mathbf r$ be reference audio and $\mathbf x=(x_1,\ldots,x_{L_x})$ the target lyric-token sequence. A frozen melody tokenizer maps $\mathbf r$ to $\mathbf m=(m_1,\ldots,m_{L_m})$ at approximately 6.25~Hz using a 512-entry codebook; it is trained from chroma-derived inputs so that the control emphasizes pitch and timing rather than lexical content. A frozen CAM++ encoder \cite{CAMPlusPlus} extracts a 192-D singer vector $\mathbf e$. A frozen Ming-UniAudio AudioVAE \cite{MingUniAudio} represents 16-kHz target audio as $\mathbf z^*\in\mathbb R^{T\times64}$ at 50~Hz.

We group $P=5$ frames into $\bar{\mathbf z}^*_k\in\mathbb R^{5\times64}$, so the acoustic recurrence runs at 10~Hz. A final short group is zero padded and retained as one patch in loss and decoding; no frame-level trimming is applied, giving at most 100-ms endpoint granularity, and this tail remains in every metric and listening item. Melody and acoustic tokens are \emph{not} one-to-one aligned: the melody prefix remains in the planner cache while semantic feedback marks generated progress. The stop head learns output patch count while target/source syllable counts may differ under reference timing cues. Inference is bounded by $\min(K_{\rm ref}+8,256)$ patches only as a rollout safeguard.

\subsection{Training Tuples and the Content Conflict}

Each tuple $(\mathbf r,\mathbf x,\mathbf z^*)$ comes from one recording: reference and target audio are the same performance, and $\mathbf x$ is its transcript. Thus training uses content-consistent reconstruction, never unavailable recordings of one melody sung with counterfactual lyrics. At evaluation, $\mathbf x$ changes while reference-derived controls remain tied to the reference audio. This paired-edit-free training turns lyric editing into a zero-paired-edit generalization test: success requires following revised lyrics despite residual source-lyric-correlated evidence, rather than fitting edit templates.

SCT/PSCG make that generalization learnable: target lyrics remain explicit, melody spans are perturbed, and an imperfect previous latent patch prevents unconditional predecessor copying. Empirically, output phones across all four operations are scored against both target and source under one max-length NED denominator. Their target-preference margin tests the realized shortcut directly rather than treating tokenizer purity as an assumption.

\subsection{State--Control--Transition Routing}

Figure~\ref{fig:sct_overview} summarizes the SCT routing. Let $\mathbf R_k$ denote the recurrent context before patch $k$:
\begin{equation}
\begin{aligned}
\mathbf R_k&=(\mathbf K^{\rm plan}_k,\mathbf K^{\rm sem}_k,\bar{\mathbf z}_{k-1}),\\
p_\theta(\bar{\mathbf z}_{1:K}\mid\mathbf x,\mathbf m,\mathbf e)
&=\prod_{k=1}^{K}p_\theta(\bar{\mathbf z}_k\mid\mathbf R_k,\mathbf x,\mathbf m,\mathbf e).
\end{aligned}
\end{equation}
The update $\mathcal U(\mathbf R_k,\bar{\mathbf z}_k)$ appends the new patch to the semantic encoder, pools its five states into semantic feedback, appends that feedback to the planner cache, and exposes the patch as the previous latent patch at $k+1$. At $k=1$, both recurrent paths are initialized with zeros.

Lyric and melody tokens prefill a Qwen3-0.6B causal planner \cite{Qwen3}. After the first step, only semantic feedback $\mathbf a_{k-1}$ is appended:
\begin{equation}
(\mathbf h_k,\mathbf K^{\rm plan}_k)
=\mathcal P_\theta(\mathbf a_{k-1},\mathbf K^{\rm plan}_{k-1}).
\end{equation}
The cache retains the full controls and all earlier feedback. A separate linear head predicts continue/stop from $\mathbf h_k$. Singer identity bypasses the planner and modulates each Flow-DiT block through adaptive layer normalization.

\paragraph{Control serialization and time bases.}
Lyrics and melody occupy distinct cached spans rather than a fused pseudo-score. Melody tokens retain reference-frame timestamps, whereas target lyric tokens retain sequence order without requiring note, duration, or edit-boundary labels. A query for patch $k$ can attend to both complete control spans and semantic feedback through $k-1$, but never to a future generated patch. AudioVAE frames use a 20-ms stride and five consecutive frames form one 100-ms transition. If the final transition is short, it is zero padded and retained as a complete patch through decoding. This contract keeps score-annotation-free controls globally available while making generated acoustic evidence strictly causal.

A frozen Whisper-style 32-layer, 1280-D causal audio encoder \cite{Whisper} provides the compact path; its states are projected to 1024-D before five-frame pooling:
\begin{equation}
\begin{aligned}
(\mathbf U_{k-1},\mathbf K^{\rm sem}_k)
&=\mathcal E_{\rm sem}(\bar{\mathbf z}_{k-1},\mathbf K^{\rm sem}_{k-1}),\\
\mathbf a_{k-1}&=P^{-1}\sum_{p=1}^{P}{\rm Proj}(\mathbf U_{k-1,p}).
\end{aligned}
\end{equation}
In parallel, the previous latent patch enters an 8-block Flow-DiT that combines conditional flow matching with a transformer denoiser \cite{FlowMatching,DiT}. Each transition receives one flow-time/planner token, five predecessor tokens, and five noisy current-patch tokens. The two paths are complementary views of the same realized patch: semantic feedback supports long-range progress, whereas the previous latent patch preserves local continuity.

The paths share each generated patch but differ in reach: semantic feedback persists in the planner cache for all later steps, whereas the previous latent patch conditions only the next Flow-DiT transition. This asymmetric lifetime separates global planning memory from local acoustic continuity.

The semantic route compresses each five-frame patch into one 1024-D feedback vector, while the local route retains the complete $5\times64$ patch for only the following transition. This design lets the planner accumulate content progress without an unrestricted acoustic history, preserves causal local continuity in Flow-DiT, and uses incremental caches to avoid re-encoding the generated prefix.

\begin{figure*}[t]
    \centering
    \includegraphics[width=\textwidth]{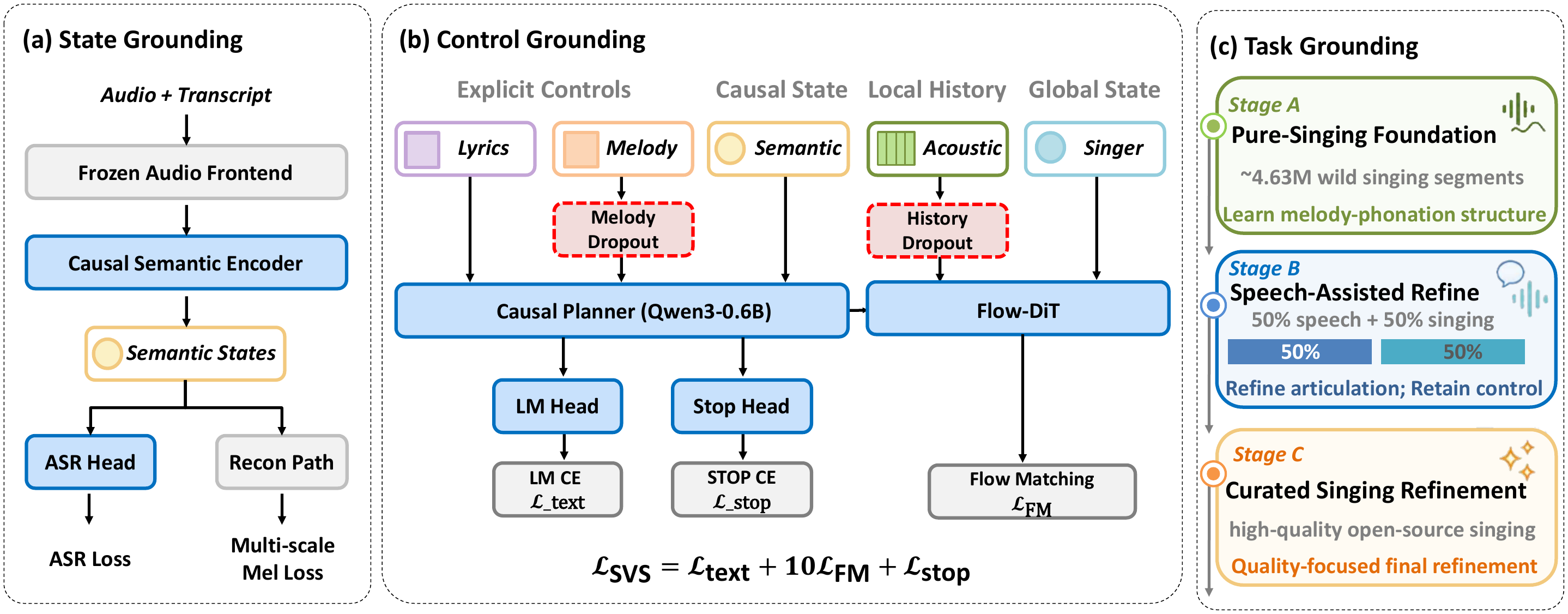}
    \caption{Progressive State--Control Grounding (PSCG). State Grounding pretrains the semantic encoder for phonetic progress; Control Grounding applies history dropout to the previous latent patch and masks melody while retaining lyrics; Task Grounding moves from broad to curated Mandarin singing.}
    \label{fig:pscg}
\end{figure*}

\subsection{Progressive State--Control Grounding}

PSCG grounds the SCT editing contract (Figure~\ref{fig:pscg}). \emph{State Grounding} pretrains the causal semantic encoder with $\mathcal L_{\rm state}=\mathcal L_{\rm ASR}+5\mathcal L_{\rm MSMel}$ before freezing it, so semantic feedback tracks realized content rather than copying acoustics. \emph{Control Grounding} applies history dropout ($p_h=.10$) to the teacher-forced previous latent patch and masks reference-melody spans ($p_m=.15$), forcing recovery from imperfect local evidence while retaining both controls. \emph{Task Grounding} schedules 72K broad-singing, 24K speech-assisted, and 24K curated-singing updates, supplying coverage before editing fidelity \cite{geng2026bridging}.

Each stage ablation disables only its defining operation while preserving width, 120K updates, optimizer, Mandarin pool, manifest, and checkpoint rule. Without State Grounding, the semantic encoder is randomly initialized and trained jointly for all 120K updates; no untrained encoder is frozen. Removed inputs are zeroed. The matched unified-context AR projects semantic feedback and the previous latent patch to the planner width, combines them through one shared recurrent route visible to both causal planning and local generation, and otherwise retains the same controls, AudioVAE, planner, Flow-DiT, data, parameters, and updates. Pretrained-frozen/pretrained-joint/random-joint/random-frozen PER is .0376/.0391/.0511/.0918. History-dropout, melody-mask, joint-corruption, and curriculum-order removals yield PER .0441/.0419/.0468/.0415; full tables are in the supplement.

For target patch $\bar{\mathbf z}^*_k$, noise $\boldsymbol\epsilon_k\sim\mathcal N(0,I)$, and $\tau\sim\mathcal U(0,1)$, conditional flow matching uses $\mathbf y_{k,\tau}=(1-\tau)\boldsymbol\epsilon_k+\tau\bar{\mathbf z}^*_k$ and target velocity $\mathbf v^*_k=\bar{\mathbf z}^*_k-\boldsymbol\epsilon_k$. The complete objective is
\begin{equation}
\mathcal L=\mathcal L_{\rm text}+10\mathcal L_{\rm FM}+\mathcal L_{\rm stop}.
\end{equation}
$\mathcal L_{\rm text}$ is causally shifted cross-entropy on nonpadding target-lyric positions during prefix teacher forcing; melody tokens, separators, and recurrent semantic feedback are excluded. It is a planner-side target-content regularizer, not direct acoustic or counterfactual supervision; its effect on generated audio is tested by ablation. $\mathcal L_{\rm stop}$ is balanced cross-entropy with the final valid patch labeled stop. Joint lyric--melody--singer dropout ($p=.10$) trains the unconditional guidance branch \cite{ClassifierFreeGuidance}, which supports classifier-free guidance during sampling. We optimize the planner, adapters, Flow-DiT, and heads while freezing all frontends. Each patch starts from Gaussian noise and uses 24 Euler steps with guidance 2.0. We emit one candidate, stop at posterior $.5$ or the rollout budget, concatenate the patches, and decode once; the companion archive includes executable reference code and module cards.

\section{Experiments}

\subsection{Data, Comparators, and Statistics}

Training audio comprises about 8,000 hours of permission-documented Mandarin singing, 2,000 hours of Mandarin Emilia speech \cite{Emilia}, and 300 hours of curated open-source singing; all data are Mandarin. Study-specific evaluation recordings came from consenting singers and contain no private information; listeners consented, and responses were de-identified. Three 120K-update seeds use BF16, ZeRO-2, AdamW, and global batch 128. The checkpoint has 0.765B trainable/2.115B total parameters, including the frozen 1.350B AudioVAE; versioned feature extractors are excluded. Cold-start/steady-tail RTF and memory cover the operational pipeline; external counts follow publisher scope.

\begin{table*}[t]
\centering
\small
\begingroup
\setlength{\tabcolsep}{3.0pt}
\begin{tabular*}{\textwidth}{@{\extracolsep{\fill}}lccccl@{}}
\toprule
System & Representation / generation & Count / basis & Required control & Quality inference & Status\\
\midrule
Vevo2 & Discrete / AR & 1.97B / publisher & Ref.\ audio + target lyrics & Official recipe & Released checkpoint\\
YingMusic-Singer-Plus & Continuous / NAR & 0.73B / publisher & Ref.\ audio + target lyrics & Official recipe & Released checkpoint\\
\textbf{\methodname} & Continuous / AR & 0.765B / 2.115B & Ref.\ audio + target lyrics & 24 steps; CFG 2.0 & Ours\\
\bottomrule
\end{tabular*}
\endgroup
\caption{Released comparators and \methodname{} share the automatic-input protocol and emit one candidate. \methodname{} count is trainable/total generative-stack size, excluding separately versioned frozen feature extractors; external counts retain publisher scope. Quality settings appear here and runtime settings in Table~\ref{tab:objective_results}.}
\label{tab:baseline_protocol}
\end{table*}

Vevo2 is the primary discrete-AR comparator: both systems derive melody automatically and decode sequentially under the common automatic-input protocol, but Vevo2 predicts discrete acoustic tokens and \methodname{} predicts continuous AudioVAE latent patches. YingMusic+ is the complementary continuous-NAR comparator with parallel inference. These comparisons establish operating profiles; causal claims rely on matched SCT/PSCG tests in Table~\ref{tab:pscg_components}.

CLA-LyricEdit-320 contains 320 edits from 160 excerpts spanning 80 songs and 40 singers, balanced across partial-line substitution (PSub), full-line substitution (FSub), deletion, and insertion. PSub replaces a contiguous 2--6-syllable span while retaining context, FSub replaces every lexical syllable in a line, and deletion and insertion remove or add 2--6 syllables. Two annotators verify targets and operations ($\kappa=.93$); all splits are song-disjoint.

FireRedASR-AED-L transcribes each output; pypinyin yields phones for PER and normalized edit distance (NED). Source-lyric reversion rate (SrcRev) counts outputs closer to source than target under NED. Melody metrics are common-voiced RMVPE log-F0 Pearson correlation (FPC) \cite{RMVPE}, voiced-onset MAE, and reference-duration deviation (RefDur; DUR-A/R1); singer similarity (SIM) is WavLM-base-plus-SV cosine \cite{WavLM}. A pre-generation stratified draw yields 80 blinded audio outputs spanning five conditions, evaluated by twenty native Mandarin listeners for naturalness (N-MOS), melody (M-MOS), and lyric intelligibility (L-MOS). We report operation-balanced macro-PER, 10,000 song-clustered draws, listener/item random effects \cite{MOSPit,MOSClustered}, and Holm correction; evaluator and listener details appear in the supplement.

\begin{center}
    \centering
    \includegraphics[width=\columnwidth]{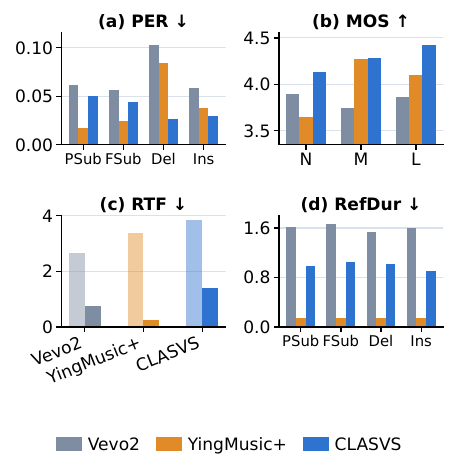}
    \captionof{figure}{Common automatic-input comparison across all three systems: operation PER, listener ratings, cold-start/steady-tail RTF (lighter/solid), and RefDur.}
    \label{fig:direct_comparison}
\end{center}

\subsection{SCT Control Audit}

Table~\ref{tab:content_audit} exposes the lexical-change baseline under a common denominator. Output--target NED is low in every operation (.0252--.0490), whereas output--source NED closely tracks the required target--source change overall (.4050 versus .3965). All four target-preference margins are positive (.2248--.7209; overall .3685). Because chroma-derived melody tokens need not be lexically invariant, this behavioral audit asks whether outputs follow the target despite any residual reference cue. Low target NED and positive margins reject simple source copying while remaining sensitive to additional errors.

\subsection{Public-Benchmark Replication}

Table~\ref{tab:public_benchmark} uses the frozen released LyricEditBench split and evaluator \cite{YingMusicSinger}, restricted to its four Mandarin-only operations. \methodname{} improves all four operations over Vevo2 and lowers macro-PER from .0761 to .0425; against YingMusic+, it improves deletion and insertion while YingMusic+ remains stronger on PSub and FSub (.0425 versus .0443 overall). This reproduces the discrete-AR gain and operation-specific profile within the Mandarin-only scope.

\begin{table*}[t]
\centering
\small
\begingroup
\setlength{\tabcolsep}{3.0pt}
\begin{tabular*}{\textwidth}{@{\extracolsep{\fill}}lcccccc@{}}
\toprule
Evaluation & PSub & FSub & Del & Ins & Overall & Interpretation\\
\midrule
Output--target NED$\downarrow$ & .0490 & .0430 & .0252 & .0289 & .0365 & Desired content\\
Output--source NED$\uparrow$ & .3107 & .7639 & .2500 & .2954 & .4050 & Rejected reference words\\
Target--source NED & .3020 & .7600 & .2380 & .2860 & .3965 & Required lexical change\\
Target-preference margin$\uparrow$ & .2617 & .7209 & .2248 & .2665 & .3685 & Output--source minus output--target\\
\bottomrule
\end{tabular*}
\endgroup
\caption{Counterfactual content audit. NED is phone edit distance divided by the larger sequence length; target-preference margin is output--source minus output--target. No-edit output--target/source is .018/.019.}
\label{tab:content_audit}

\begingroup
\setlength{\tabcolsep}{4.0pt}
\begin{tabular*}{\textwidth}{@{\extracolsep{\fill}}lccccr@{}}
\toprule
System & PSub & FSub & Del & Ins & Macro-PER$\downarrow$\\
\midrule
Vevo2 & .0664 & .0607 & .1123 & .0649 & .0761\\
YingMusic-Singer-Plus & \textbf{.0214} & \textbf{.0186} & .0946 & .0426 & .0443\\
\textbf{\methodname} & .0508 & .0435 & \textbf{.0398} & \textbf{.0357} & \textbf{.0425}\\
\bottomrule
\end{tabular*}
\endgroup
\caption{Four Mandarin-only LyricEditBench operations under the common automatic-input protocol and public evaluator. Cross-lingual categories are excluded; the discrete-AR gain and operation-specific profile transfer.}
\label{tab:public_benchmark}
\end{table*}

\begin{table*}[t]
\centering
\small
\begingroup
\setlength{\tabcolsep}{2.4pt}
\begin{tabular*}{\textwidth}{@{\extracolsep{\fill}}lcccccccccc@{}}
\toprule
System & PSub & FSub & Del & Ins & Macro-PER [95\% CI]$\downarrow$ & FPC$\uparrow$ & SIM$\uparrow$ & RefDur./Onset$\downarrow$ & C/S RTF$\downarrow$ & $\Delta$GiB$\downarrow$\\
\midrule
Vevo2 & .0617 & .0563 & .1033 & .0581 & .0699 [.0617,.0781] & .7440 & .892 & 1.6009\%/69.58ms & \textbf{2.649}/.744 & 8.62\\
YingMusic-Singer-Plus & \textbf{.0171} & \textbf{.0245} & .0847 & .0382 & .0411 [.0356,.0469] & .9340 & .906 & \textbf{.1464\%}/18.80ms & 3.392/\textbf{.241} & 6.74\\
\textbf{\methodname} & .0505 & .0442 & \textbf{.0260} & \textbf{.0298} & \textbf{.0376 [.0326,.0429]} & \textbf{.9410} & \textbf{.914} & .9923\%/\textbf{7.68ms} & 3.840/1.395 & \textbf{5.82}\\
\bottomrule
\end{tabular*}
\endgroup
\caption{One candidate per item/run on 320 Mandarin edits. RefDur is DUR-A/R1 absolute output/reference duration deviation; Onset is voiced-onset MAE. \methodname{} intervals resample songs and three training runs; released comparators condition on one checkpoint. C/S is cold-start/steady-tail RTF; $\Delta$GiB is incremental peak memory in a separate runtime benchmark.}
\label{tab:objective_results}
\end{table*}

\begin{table}[t]
\centering
\small
\begingroup
\setlength{\tabcolsep}{3.1pt}
\begin{tabular*}{\columnwidth}{@{\extracolsep{\fill}}lccc@{}}
\toprule
Configuration & PER$\downarrow$ & FPC$\uparrow$ & SrcRev$\downarrow$\\
\midrule
\textbf{Full SCT + PSCG} & \textbf{.0376} & \textbf{.9410} & \textbf{.081}\\
No semantic-feedback route & .0564 & .9205 & .154\\
No latent-patch route & .0492 & .8896 & .086\\
Unified-context AR & .0447 & .9279 & .118\\
\midrule
$-$ State Grounding & .0511 & .9168 & .143\\
$-$ Control Grounding & .0468 & .9242 & .121\\
$-$ Task Grounding & .0439 & .9298 & .102\\
\bottomrule
\end{tabular*}
\endgroup
\caption{SCT and PSCG checks under one locked continuous-AR protocol. Unified-context AR is matched in frontends, planner, generator, data, width, parameters, and updates; lower SrcRev is better.}
\label{tab:pscg_components}
\end{table}

\begin{table}[t]
\centering
\small
\begingroup
\setlength{\tabcolsep}{2.0pt}
\begin{tabular*}{\columnwidth}{@{\extracolsep{\fill}}lccc@{}}
\toprule
System & N-MOS & M-MOS & L-MOS\\
\midrule
Vevo2 & \shortstack{3.89\\{[3.78,4.00]}} & \shortstack{3.74\\{[3.63,3.85]}} & \shortstack{3.86\\{[3.75,3.97]}}\\
YingMusic+ & \shortstack{3.65\\{[3.55,3.74]}} & \shortstack{4.27\\{[4.19,4.35]}} & \shortstack{4.10\\{[4.01,4.19]}}\\
\textbf{\methodname} & \shortstack{\textbf{4.13}\\\textbf{[4.04,4.22]}} & \shortstack{4.28\\{[4.20,4.36]}} & \shortstack{\textbf{4.42}\\\textbf{[4.34,4.50]}}\\
\midrule
$\Delta$ vs.\ YingMusic+ & +.48 & +.01 & +.32\\
Adjusted $p$ & $<.001$ & .79 & $<.001$\\
\bottomrule
\end{tabular*}
\endgroup
\caption{Ordinal listener--item analysis over 80 blinded audio outputs; deltas use YingMusic+, and the M-MOS difference is not detectable.}
\label{tab:mos_results}
\end{table}

\subsection{Editing Accuracy, Quality, and Identity}

Table~\ref{tab:objective_results} and Figure~\ref{fig:direct_comparison} show a broad gain over the discrete-AR comparator: relative to Vevo2, \methodname{} reduces macro-PER by 46.2\%, improves every operation, FPC, SIM, RefDur, voiced-onset MAE, all three MOS scales, and peak memory, at the cost of slower RTF. Against the continuous-NAR YingMusic+ comparator, \methodname{} has lower macro-PER by $-.0035$ [$-.0056,-.0013$] and better deletion/insertion accuracy (.0279 versus .0615 mean). \methodname{} N-MOS rises from 3.65 to 4.13 ($+.48$ [$+.34,+.62$], adjusted $p<.001$), reflecting a marked listener preference in naturalness and acoustic clarity. YingMusic+ remains stronger on substitution, global RefDur, and steady-tail RTF, while Vevo2 has the lowest cold-start RTF. The comparison therefore supports a complete-system continuous-AR operating point rather than uniform or one-factor architectural dominance.

\subsection{PSCG and Inference Diagnostics}

\looseness=-1
PSCG development sweeps favor moderate latent-patch corruption: increasing $p_h$ from .10 to .20 lowers PSub PER from .0505 to .0437 and SrcRev from .081 to .055, but worsens macro-PER to .0397; .10 is the locked trade-off. Removing State Grounding changes PER/FPC to $.0511/.9168$, supporting semantic feedback as an explicitly grounded~phonetic-progress~State.

Table~\ref{tab:pscg_components} uses three seeds. Full versus matched unified improves PER by $-.0071$ [$-.0085,-.0057$], FPC by $+.0131$ [$+.0098,+.0164$], and SrcRev by $-.037$ [$-.048,-.026$]. The route removals are complementary rather than interchangeable: removing semantic feedback most strongly harms lyric accuracy and target following, whereas removing the previous latent patch most strongly harms melody preservation. Grounding-stage removals show the same division of labor at training time. Removing $\mathcal L_{\rm text}$ raises PER to .0458 and SrcRev to .137, further connecting target organization in the causal planner to generated audio.

\paragraph{Configuration discipline.}
All matched rows retain the same frontends, data, width, 0.765B trainable budget, 120K updates, manifest, and checkpoint rule; three-seed PER SD across routing/stage rows is .0009--.0016. The explicit $2{\times}2$ State-Grounding check separates initialization from freezing; its one-time semantic pretraining is an additional cost outside the downstream schedule, so this check is not compute matched. Separate removals isolate history dropout, melody masking, and curriculum order. Released checkpoints remain system comparators, not causal ablations; runtime settings are separately labeled.

\begin{table*}[t]
    \centering
    \includegraphics[width=.82\textwidth]{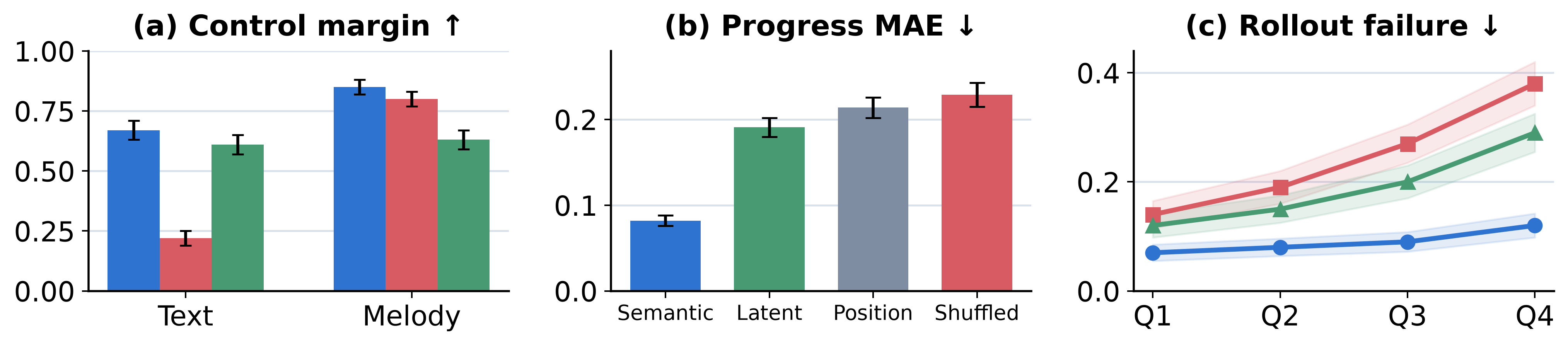}
\captionof{figure}{SCT diagnostics (song-clustered 95\% intervals): paired control swaps, a progress probe, and failure by output-length quartile. Blue/red/green denote full/semantic-feedback-shuffled/latent-patch-shuffled; gray is position-only.}
    \label{fig:control_rollout}
\end{table*}

\paragraph{Inference-time sensitivity checks.}
Figure~\ref{fig:control_rollout} creates held-out counterfactual conflicts. Target-lyric swaps fix the reference-melody control, singer vector, flow noise, and rollout budget; matched reference-melody swaps fix all other inputs. Order-breaking and replacement tests agree: disrupting semantic feedback lowers the target-lyric control margin, disrupting latent patches lowers the reference-melody control margin, and clustered intervals exclude zero. A progress probe favors semantic feedback.

\section{Analysis and Discussion}

\paragraph{System-level evidence.}
Vevo2 and YingMusic+ are the discrete-AR and continuous-NAR comparators under the common automatic-input protocol. They establish complete-system profiles; matched unified-context, route, and stage interventions support SCT/PSCG attribution. The public Mandarin benchmark reproduces the operation-specific profile.

\paragraph{Why the two comparators are complementary.}
Vevo2 shares \methodname{}'s sequential, automatically extracted melody-token interface but predicts discrete acoustic tokens instead of continuous AudioVAE latent patches; \methodname{} improves every reported editing and preservation measure. YingMusic+ leads substitution and steady-tail RTF; Vevo2 leads cold-start RTF; \methodname{} leads deletion, insertion, voiced-onset MAE, N-MOS, L-MOS, and macro-PER. Together with the matched unified-context ablation, these complementary profiles support---without factorial attribution---high-fidelity stepwise control with learned stopping.

\paragraph{Zero-paired-edit generalization.}
Resinging revised words changes the performance, so paired counterfactual edit recordings are not naturally observable. Training never reveals the conflict, edit-span alignment, or operation template; \methodname{} applies its SCT editing contract only at test time. The .3685 target-preference margin and deletion/insertion gains evidence behavior beyond reconstruction; progress-probe MAE (.082 versus .191), route shuffles, and history dropout independently support the contract across held-out cases.

\paragraph{Robustness across edit demands.}
\methodname{} improves every Vevo2 operation, especially length-changing edits. Against YingMusic+ it leads deletion/insertion but trails substitution; the same profile recurs on the public benchmark under matched evaluation. Output--target/source NED rules out retained-word shortcuts; RefDur and voiced-onset MAE track reference pacing. Thus lexical, temporal, and replicated evidence complement macro-PER.

\paragraph{State size and runtime.}
Here \emph{compact} is temporal: one 1024-D semantic-feedback vector enters the cache per patch; the previous latent patch stays one patch wide. Cache growth remains linear during long rollouts and incremental memory is 5.82~GiB; long inputs require future compression.

\paragraph{Metric and deployment trade-offs.}
\methodname{} improves Vevo2's editing and preservation measures under matched inputs, at slower RTF. Against YingMusic+ it leads macro-PER, deletion/insertion, naturalness, and voiced-onset MAE but trails substitution, global RefDur, and RTF. With cold-start/steady-tail RTF 3.840/1.395, it remains better suited to offline use.

\paragraph{What the stopping audit adds.}
The stop head determines realized patch count across all operations; $\min(K_{\rm ref}+8,256)$ is only a guard. Budget-hit rates are 0/0/0/1.25\% for PSub/FSub/Del/Ins; premature/delayed-stop rates never exceed 1.25/2.50\%. Here ``length change'' refers to target/source syllable count, not a required change in global audio duration: insertion and deletion may redistribute local rate while retaining the reference timeline. Together with output--source NED, these rates show that deletion accuracy is not obtained by truncation or retained words.

\paragraph{Transfer audits.}
Across 24 post-freeze singers (192 edits), \methodname{}/YingMusic+ gives PER .0472/.0521, FPC .934/.928, SIM .905/.898, and voiced-onset MAE 9.4/21.6~ms. The public benchmark retains YingMusic+'s substitution lead and \methodname{}'s deletion/insertion advantage, confirming the operation-specific profile.

\paragraph{Perceptual interpretation.}
\methodname{} improves N-MOS by $+.48$ [$+.34,+.62$] and L-MOS by $+.32$ [$+.20,+.44$] over YingMusic+ under the same listening protocol; M-MOS differs by an undetectable $+.01$. N-MOS explicitly includes clarity and artifacts, so 4.13 versus 3.65 supports clearer output; L-MOS independently confirms target intelligibility. These perceptual gains complement deletion/insertion accuracy without treating the M-MOS null as equivalence.

\section{Limitations, Reproducibility, and Ethics}

Evaluation uses one recognizer and reference-timing metrics on Mandarin 2--6-syllable edits; cross-lingual and long-edit validation remain open. Upon publication, we will release training, inference, and evaluation code/configurations, per-item scores, CLA-LyricEdit-320 annotations and permitted audio, plus identity-authorized inference checkpoints; restricted training audio cannot be redistributed.

Recognizable-voice editing can enable impersonation \cite{CtrSVDD}. Source-level permissions, opt-out and removal procedures, identity-authorized access, and localized provenance therefore govern data use, checkpoint release, and responses to misuse reports \cite{AudioSeal}.

\section{Conclusion}

\methodname{} establishes continuous-latent autoregression for melody-preserving Mandarin lyric editing without manual score annotations. Its SCT editing contract separates target-lyric and reference-melody controls, semantic feedback on phonetic progress, and the previous AudioVAE latent patch; PSCG grounds this routing through content-consistent reconstruction. It follows revised lyrics without paired counterfactual edit recordings. Under the common automatic-input protocol, \methodname{} improves every Vevo2 operation and attains the lowest macro-PER, especially on deletion and insertion. It preserves melody and singer identity while improving naturalness and lyric intelligibility over the continuous-NAR comparator, though parallel generation is faster and stronger on substitution. Interventions show semantic feedback guides progress while the previous latent patch preserves local continuity. These results provide a continuous-AR foundation for broader controllable singing performance.

\bibliography{arxiv2027}

\end{document}